\documentclass[preprint,aps,amsmath,pre,sort&compress]{elsarticle}
\usepackage{graphicx}
\usepackage{amsmath}
\usepackage{caption}
\usepackage{subcaption}
\usepackage{color}
\usepackage{lineno}

\begin{document}

\title{Registering the evolutionary history in individual-based models of speciation}

\author[unicamp_b]{Carolina L. N. Costa}\corref{cor1}
\ead{lemes.carol@gmail.com}
\author[unicamp_f]{Flavia M. D. Marquitti}
\author[la_plata,unicamp_f]{S. Ivan Perez}
\author[unicamp_f]{David M. Schneider}
\author[unicamp_f]{Marlon F. Ramos}
\author[unicamp_b,unicamp_f]{Marcus A.M. de Aguiar}

\address[unicamp_b]{Instituto de Biologia,
Universidade Estadual de Campinas, Unicamp, 13083-859, Campinas, SP, Brazil}

\address[unicamp_f]{Instituto de F\'{\i}sica `Gleb Wataghin',
Universidade Estadual de Campinas, Unicamp, 13083-859, Campinas, SP, Brazil}

\address[la_plata]{Divisi\'on Antropolog\'{\i}a, Museo de La Plata, Universidad Nacional de La Plata, Paseo del Bosque s/n, 1900 La Plata, Argentina}

\cortext[cor1]{Corresponding author. Phone: +55-19-35215466}

\begin{abstract}

Understanding the emergence of biodiversity patterns in nature is a central problem in biology.
Theoretical models of speciation have addressed this question in the macroecological scale, but
little has been investigated in the macroevolutionary context. Knowledge of the evolutionary
history allows the study of patterns underlying the processes considered in these
models, revealing their signatures and the role of speciation and extinction in shaping macroevolutionary patterns. In this
paper we introduce two algorithms to record the evolutionary history of populations in
individual-based models of speciation, from which genealogies and phylogenies can be constructed.
The first algorithm relies on saving ancestral-descendant relationships, generating a matrix that
contains the times to the most recent common ancestor between all pairs of individuals at every
generation (the Most Recent Common Ancestor Time matrix, MRCAT). The second algorithm directly
records all speciation and extinction events throughout the evolutionary process, generating a
matrix with the true phylogeny of species (the Sequential Speciation and Extinction Events, SSEE).
We illustrate the use of these algorithms in a spatially explicit individual-based model of
speciation. We compare the trees generated via MRCAT and SSEE algorithms with trees inferred by
methods that use only genetic distance among extant species, commonly used in empirical studies and
applied here to simulated genetic data. Comparisons between tress are performed with metrics
describing the overall topology, branch length distribution and imbalance of trees. We observe that
both MRCAT and distance-based trees differ from the true phylogeny, with the first being closer to
the true tree than the second.
 

\end{abstract}

\maketitle

\section{Introduction}
\label{intro}

The origin of the patterns of diversity at macroecological scale is a central problem in biology
\cite{coyne2004speciation,de2009global,gavrilets2014models}. In the last decades these
macroecological patterns, such as geographical variation in species richness, species abundance
distributions and species-area relationships, have been studied from empirical and theoretical
perspectives \cite{turelli2001theory,field2009spatial,martins2013evolution,May20141657,Kopp2010}. In
the theoretical context, neutral models of speciation –- where differences between individuals are
irrelevant for their birth, death, and dispersal rates \cite{gavrilets2014models,hubell2001unt} –-
have played a central role in understanding the patterns of diversity at the macroecological
scale. With the help of computers, it became possible to test different hypothesis about mechanisms
that might drive speciation, such as sympatric versus allopatric processes, assortative mating and
the effect of number of genes \cite{gavrilets2000patterns,Dieckmann1999,rettelbach2013three}.

Among the different approaches designed to quantitatively study speciation
\cite{gavrilets2014models,gavrilets2003perspective}, models that explicitly incorporate space have
allowed the study of major macroecological patterns that could be compared with those observed in nature
\cite{de2009global,May20141657,Gomes2012,GarciaMartin05072006}. However, these models have given
little attention to the historical or evolutionary dimension of the origin of diversity, which is reflected
in the macroevolutionary patterns described by phylogenetic trees
\cite{manceau2015phylogenies,pigot2010shape,hagen2015age,quental2011molecular}. Because of the
increased interest in the role of microevolutionary processes on the resulting macroecological
patterns, the extension of these approaches for keeping track the branching or phylogenetic
divergence process is a next fundamental step to further explore models of speciation using
simulations \cite{manceau2015phylogenies,davies2011neutral,rosindell2015unifying}. Individual-based
models (IBM) widely used in biology \cite{deangelis2014individual} have the advantage that can be
easily extended to include this historical perspective and to provide a record of the
ancestral-descendent relationships among the simulated individuals and/or species. These
relationships can be stored in matrices from which individual genealogies and species trees (i.e.
phylogenies) may be directly obtained.

In this article we describe two algorithms to save historical information in individual-based models
of speciation. The first algorithm focus on genealogies and the quantity saved is the parenthood of
each individual. With parenthood registered, the {\it time to the most recent common ancestor},
i.e., the number of generations needed to go backward to find a common ancestor of one individual with
another individual of the population, can be easily calculated in terms of the common ancestral of
the parents. These times are computed at every generation between all pairs of individuals and, in
the end of the simulation, are saved in a matrix (the Most Recent Common Ancestor Time matrix -
MRCAT). The second algorithm focus on phylogenies and consists in directly record all speciation and
extinction events (the Sequential Speciation and Extinction Events - SSEE) and setup a matrix
analogous to MRCAT but whose entries are species rather than individuals. The SSEE matrix contains
the exact branching times in the simulated clade or community, including all extinct species. The
MRCAT and SSEE matrices can be used to drawn the exact branching sequence of the simulated
individuals and species, respectively. This procedure differs from the inference methods based on
phenotypic and genetic traits used to estimate phylogenies in natural studies, because in our model 
we are looking for the branching process forward in time, while in usual approaches the same process 
is looked backwards in time. In addition to the presentation of the MRCAT and SSEE algorithms, we compare 
the trees they generate with those obtained by usual distance-based methods of phylogenetic inference using only genetic
data from simulated individuals of the final species. Comparing these inferred phylogenies with
those generated by MRCAT or SSEE algorithms might offer a practical way to evaluate the reliability
of the estimated trees to recover natural macroevolutionary patterns.

The paper is organized as follows: in section 2 we describe the algorithms to record
ancestor-descendant relationships (MRCAT, subsection 2.1) and speciation/extinction events (SSEE,
subsection 2.2). In section 2.3 we compare the true phylogenetic tree drawn from the SSEE algorithm
with genealogies of individuals drawn in the MRCAT algorithm considering only one individual per
species. In section 3 we discuss the applications of the algorithms proposed in section 2. First, we
present an individual-based model of speciation proposed in \cite{de2009global} in which the
algorithms regarding the ancestor-descendant relationships and the branching process were
incorporated (subsection 3.1). We emphasize that the algorithms are quite general and could be
implemented in most IBM's. Next, we briefly describe the Unweighted Paired Group Method with
Arithmetic mean (UPGMA) \cite{murtagh1984complexities}, the Neighbor Joining (NJ)
\cite{saitou1987neighbor} and the Minimum Evolution (ME) \cite{rzhetsky1993theoretical} methods,
which are based on genetic distances calculated directly from one individual of each species present
in the last generation of the simulation (subsection 3.2). While closer to what empiricists do, the
phylogenies derived from these methods are more distant from the true phylogeny generated by the
SSEE algorithm than the phylogeny based on the MRCAT algorithm presented here. We end this section
presenting the statistical measurements utilized to compare phylogenies obtained from algorithms
proposed here with those estimated by distance-based methods (subsection 3.3). The goal is to show
that the accuracy of some methods usually employed when the only information available is the data
of individuals collected from nature can be evaluated with the help of models. In section 4 we
present the results regarding the output of simulations and the comparisons of phylogenies. Finally,
section 5 was devoted to discussion and section 6 to conclusions.

\section{Registering the history of individuals and species}
\label{algorithms}

In this section we describe two algorithms to record historical information during the evolution of a
population. The first algorithm records genealogical relationships between all pairs of individuals
at every generation. The second, in turn, registers all the speciation and extinction events that
occur along the evolutionary history. These algorithms are general enough to be applied to most
individual-based models of speciation.

\subsection{Ancestral-descendant relationships among individuals - MRCAT}
\label{mrcat}

In this subsection we show how the time to the most recent common ancestor between all pairs of
individuals can be obtained by keeping track of parental relationships at every generation. We also
show how this information can be used to draw the genealogy of individuals of the last simulated
generation. We distinguish between asexual and sexual models because of the technical differences in
tracking only one or two parents.

\subsubsection{Asexual models}
\label{mrcata}

Consider a population of $N_t$ asexual individuals at generation $t$. At the end of generation $t$ a list of 
the new individuals comprising generation $t+1$, along with a list of their respective parents is produced. 
An example is shown in Table \ref{table1}.

 \begin{table}[]
	\centering
	\begin{tabular}{c | c }
		Individuals at generation $t+1$ 	    &  Parent at generation $t$ \\
		\hline \hline
		1	  &  $P(1) = 4$  \\
		2   &  $P(2) = 8$  \\
		3	  &  $P(3) = 1$  \\
		4   &  $P(4) = 4$  \\
		\dots  &  \dots \\
		$N_{t+1}$  &  $P(N_{t+1}) = 15$ 
	\end{tabular} 
	\caption{List of individuals ($i$) at generation $t+1$ and their respective parents ($P(i)$) at
		generation $t$ in an asexual model. This information is necessary to construct the MRCAT matrix.
		Parents of each individual must be recorded to track the most recent common ancestor between
		individuals at the end of a simulation. Note that individuals at generation $t$ are not the same
		individuals at generation $t+1$ (discrete generations).\\} 
	\label{table1}
\end{table}

\begin{figure}
	 \centering \includegraphics[width=1.0\linewidth]{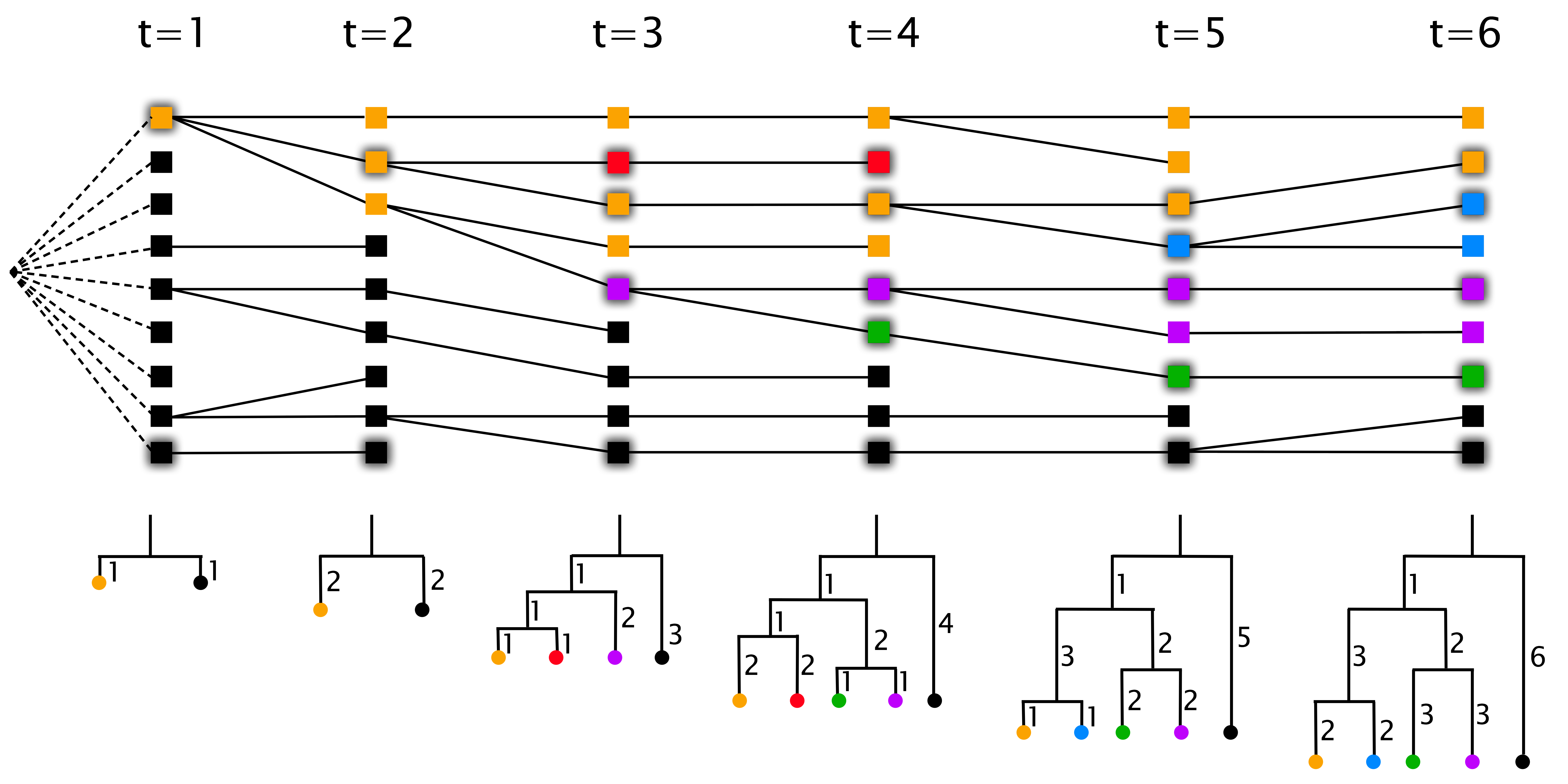} 
	\caption{Illustration of ancestor-descendant relationships for an asexual population with constant size
		$N=9$ implemented with MRCAT algorithm. Each square is an individual and colors represent different species. 
		Phylogenetic trees are constructed by selecting one individual per species (shaded squares). } 
	\label{figMRCAT}
\end{figure}

The parent of individual $i$ in generation $t+1$ is denoted $P(i)$. In the example in Table
\ref{table1}, $P(1)=4$, $P(2)=8$, $P(3)=1$, etc. The MRCAT between individuals $i$ and $j$ is,
therefore, 
\begin{equation}
T_{t+1}(i,j) = T_t(P(i),P(j)) + 1.
\label{mrcatasexual}
\end{equation}
which is simply the time to the most recent common ancestor between the parents plus one, since a
generation has passed \cite{higgs1992genetic}. As examples
\begin{displaymath}
T_{t+1}(1,2) = T_t(4,8) + 1 
\end{displaymath}
and
\begin{displaymath}
T_{t+1}(1,4) = T_t(4,4) + 1 = 1.
\end{displaymath}
since in this last case they have the same parent. Starting from $T_0(i,j)=1$ if $i\neq j$ and noting
that $T_t(i,i) = 0$ at all times the rule (\ref{mrcatasexual}) allows one to compute the MRCAT
matrix for any number of generations. The matrix $T$ is stored only for two times, the past and the
present generation, so that the memory cost does not depend on time, only on the (square) size of
population. A schematic view of the algorithm is shown in Fig. \ref{figMRCAT}, where the
genealogical relationships between 9 individuals originated from a single ancestral is represented. In this
example the total population size is kept fixed, so that the full MRCAT matrix is always $9 \times 9$. 
The phylogeny of the community can be drawn by selecting one individual per species at each moment
in time. The corresponding matrices at $t=3$ and $t=6$ are given by

\begin{equation}
	T_{3} =
	\left(
	\begin{array}{cccc}
		0 \; & 1 \; & 2 \; & 3 \\
		1 \; & 0 \; & 2 \; & 3 \\
		2 \; & 2 \; & 0 \; & 3 \\ 
		3 \; & 3 \; & 3 \; & 0 
	\end{array}
	\right) ;\qquad \qquad
	T_{6} =
	\left(
	\begin{array}{ccccc}
		0 \; & 2 \; & 5 \; & 5 \; & 6 \\
		2 \; & 0 \; & 5 \; & 5 \; & 6 \\
		5 \; & 5 \; & 0 \; & 3 \; & 6 \\ 
		5 \; & 5 \; & 3 \; & 0 \; & 6 \\
		6 \; & 6 \; & 6 \; & 6 \; & 0
	\end{array}
	\right).
	\label{matmrcat}
\end{equation}
where the selected individuals are shown in shaded colors (from top to bottom) at the corresponding times.

%

\subsubsection{Sexual models}
\label{mrcats}

The generation of MRCAT matrices in sexual models is slightly different, since each individual $i$
has two parents, a mother $P_1(i)$ and a father $P_2(i)$. Consider as an example a population
which has 4 females and 3 males in generation $t$ and gives rise to 5 females and 3 males in
generation $t+1$ (Table \ref{table2}). Notice that not only the total number of individuals but also
the number of males and females may vary over generations. As the model is sexual, both maternal and
paternal lineages can be followed in the simulations, allowing the generation of two different MRCAT
matrices and their corresponding trees. A third option is not tracking lineages by sex, but record
the most recent common ancestor taking into account both parents, which is the only option if the
model considers hermaphroditic individuals.\\

\begin{table}[]
	\centering
	\begin{tabular}{c | c | c }
		Individuals at generation $t+1$ &  Mother at generation $t$ &  Father at generation $t$ \\
		\hline \hline
		Females & & \\
		\hline
		1	  &  $P_1(1) = 4$ & $P_2(1) = 6$  \\
		2   &  $P_1(2) = 3$ & $P_2(2) = 7$  \\
		3	  &  $P_1(3) = 1$ & $P_2(3) = 7$  \\
		4   &   $P_1(4) = 4$ & $P_2(4) = 5$  \\
		5   &   $P_1(5) = 2$ & $P_2(5) = 6$  \\
		\hline
		Males & & \\
		 6  &   $P_1(6) = 1$ & $P_2(6) = 5$ \\
		 7  &  $P_1(7) = 3$ & $P_2(7) = 5$  \\
		 8	 &  $P_1(8) = 3$ & $P_2(8) = 7$  \\
	\end{tabular} 
	\caption{List of individuals ($i$) at generation $t+1$ and their respective parents ($P_1(i)=
	mother$ and $P_2(i)=father$) at generation $t$ in a sexual model. In this case each individual has
	two parents, $P_1$ and $P_2$. Notice that the couple 3 and 7 at generation $t$ had two offspring,
	the individuals 2 and 8 at generation $t+1$, while other couples had only one offspring.
	Additionally, notice that there were 4 females and 3 males at generation $t$, while there are 5
	females and 3 males at generation $t+1$.\\ } 
	\label{table2}
\end{table}

-- {\it Maternal and paternal lineages.} The maternal lineage of individuals is obtained by computing the time to the most recent common ancestor of their corresponding mothers: 
\begin{equation}
T^M_{t+1}(i,j) = T^M_t(P_1(i),P_1(j)) + 1
\label{mrcatmother}
\end{equation}
with $T_0^M(i,j) = 1$ if $i \neq j$ and $T^M_t(i,i) = 0$. Similarly, the paternal lineage is computed with
\begin{equation}
T^F_{t+1}(i,j) = T^F_t(P_2(i),P_2(j)) + 1
\label{mrcatfather}
\end{equation}
with $T_0^F(i,j) = 1$ if $i \neq j$ and $T^F_t(i,i) = 0$. Both $T^M$ and $T^F$ are computed for all
individuals, females and males. \\

-- {\it Lineages of hermaphroditic individuals.} Many simulations consider, for simplicity, hermaphroditic
individuals. In this case, the separation into maternal and paternal lineages does not make sense
and the definition of the MRCAT matrix is 
\begin{equation}
T_{t+1}(i,j) = \mbox{min}_{\{k,l\}} \{ T_t(P_k(i),P_l(j)) \} + 1
\label{mrcatherm}
\end{equation}
with $k,l=\{1,2\}$, $T_0(i,j) = 1$ and $T_t(i,i) = 0$. This considers, literally, the most recent
common ancestor of $i$ and $j$, taking all parental combinations into account. The same definition
is applied to sexual models with sex separation when the recorded genealogy does not separate the
maternal and paternal lineages. In the case of hermaphroditic model the MRCAT matrix does not
determine the tree uniquely. A detailed example of this situation is described in Supporting
Information, section I. 

\subsubsection{Drawing genealogies from MRCAT matrices}
\label{genealogy}

At the end of the simulated evolutionary process the MRCAT matrix contains the time to the most
recent common ancestor between every pair of individuals of the extant population and this
information can be used to draw genealogical trees. Drawing the tree from the MRCAT matrix consists
in joining individuals into groups according to their most recent common ancestral (Fig. \ref{figMRCAT}). The tree starts
with $N$ {\it units} (the extant individuals) and at each step of the process two of these units are
joined together to form a group, so that the number of units decreases by 1. Next, the time to the
most recent common ancestral between the newly formed group and the other units of the tree
(previously formed groups or extant individuals) are recalculated with a so called {\it clustering
	method}. The process ends when a single unit is left, the root of the tree. As discussed in the SI,
section I, a unique tree is generated independently of the clustering method for asexual, maternal
or paternal lineages. For hermaphroditic populations or for sex separation but with the MRCA taking
into account both parents that is not the case. In these situations more than one tree can be drawn
from the same MRCAT matrix using different clustering procedures.

In all cases the tips (or leaves) of the tree represent extant individuals whereas internal nodes
represent the most recent common ancestor between a pair of individuals. Branch length
denote the time in generations between an ancestor and its descendants (see, for instance, Fig.
S1 in the SI). The x-axis on the base of the tree display the extant individuals spaced by one unit,
if one is only interested in the tree topology and statistics of time intervals, or it might
represent the genetic distance between individuals, as shown, for instance, in subsection \ref{gdm}. More information about the drawing of trees is available in Supporting Information, section II.

\subsection{Recording all speciation and extinction events - SSEE}
\label{ssee}

\begin{figure}[!htpb]
  \centering \includegraphics[width=1.0\linewidth]{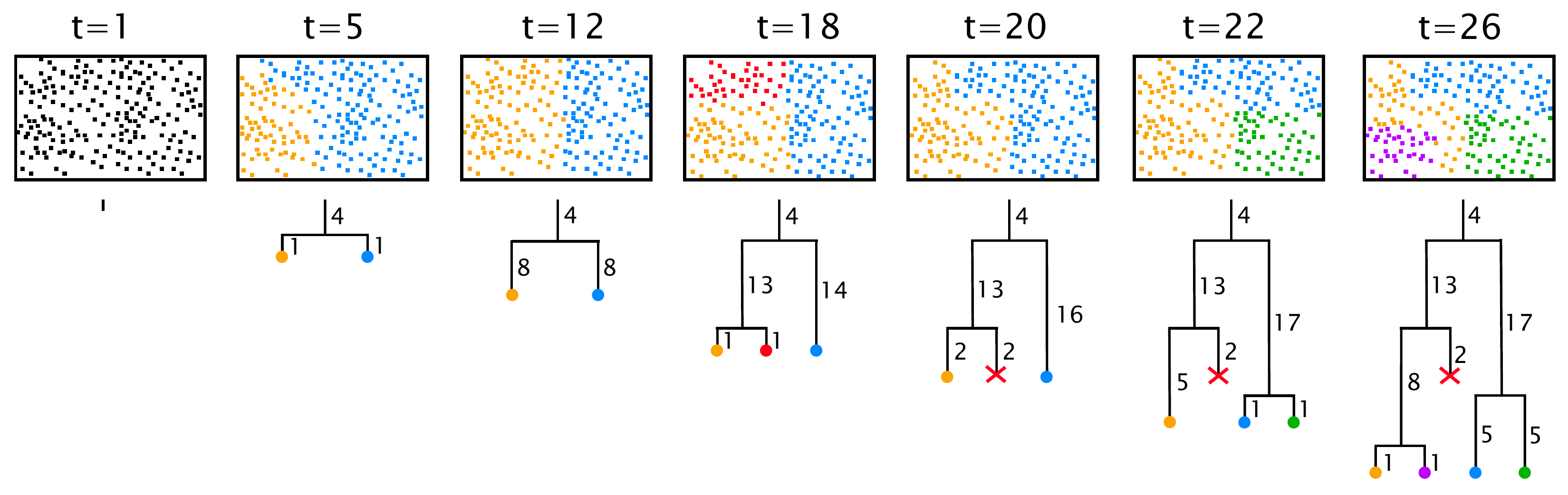}
  \caption{Illustration of speciation and extinction events implemented with SSEE algorithm and the corresponding phylogenetic trees exhibiting the complete history.}
\label{figext}
\end{figure}

The algorithm described in subsection \ref{mrcat} records the ancestral-descendant relationships
between all pairs of individuals in the population at a given point in time. This allows the drawing
of entire genealogies. However, information about individuals that died without leaving descendants
or species that went extinct is totally lost. In this subsection we describe an algorithm that
allows the drawing of the true phylogenetic tree, retaining information about all species that ever
existed during the evolution (Fig. \ref{figext}).

We will use a new matrix $S_t$ (the SSEE matrix) such that $S_t(i,j)$ is the time when species $i$
and $j$ branched off a common ancestral species. Species that go extinct will be kept in the matrix
but will be assigned a label to distinguish them from living (extant) species. This label will be
stored in a {\it extinction vector} $E_t$ such that $E_t(i)=0$ indicates a living species at time
$t$ and $E_t(i) = \tau \neq 0$ indicates the moment $\tau$ when the species disappeared.

The algorithm is as follows: consider the hypothetical sequence os speciation and extinction events
displayed in Fig. \ref{figext}. At time t=18 there are three species that we denote as Orange(18), Red(18)
and Blue(18) and the corresponding S matrix and E vector are
\begin{equation}
S_{18} =
\left(
\begin{array}{ccc}
0 \; & 1 \; & 14 \\
1 & 0 & 14 \\
14 & 14 & 0 
\end{array}
\right) ;\qquad \qquad
E_{18} =
\left(
\begin{array}{c}
0 \\
0 \\
0
\end{array}
\right).
\label{mat18e}
\end{equation}

At two steps further, $t=20$, one finds only two species, Orange(20) and Blue(20). Notice that names (and
colours) are arbitrary and to determine the relation between these species and the ones at the
previous time step we need to look at the parents of individuals in each species. Suppose, as
illustrated in the figure, that we find that the parents of individuals in Blue(20) belonged to
species Blue(18). In this case we draw a link between Blue(18) and Blue(20) and mark Blue(18) as a
species that ’survived’ that time step, i.e., we set $E_{20}(1)=0$. Similarly Orange(20) links with
Orange(18) and $E_{20}(2)=0$. Looking at the previous generation we notice that species Red(18) did not
leave any descendant species, i.e., it went extinct. In order to keep track of it we create a
virtual species Red(20) and set $E_{20}(3) = 20$ as a mark that it is no longer a living species and
went extinct at time 20. The SSEE and E vector at time 20 become 
\begin{equation}
	S_{20} =
	\left(
	\begin{array}{ccc}
		0 \; & 16 \; & 2 \\
		16 & 0 & 16 \\
		2 & 16 & 0 
	\end{array}
	\right) ;\qquad \qquad
	E_{20} =
	\left(
	\begin{array}{c}
		0 \\
		0 \\
		20
	\end{array}
	\right).
	\label{mat19e}
\end{equation}

Extinct species are, therefore, treated as species that will never again speciate, but will be kept
in the matrix. When drawing the corresponding tree its branch will stop at the value $E(i)$.
Proceeding in this way, with the living species always filling the first part of the matrix,
followed by copies of extinct species, we can draw the complete phylogeny and study extinction
dynamics as well. At time $t=26$ the SSEE matrix and extinction vector E are
\begin{equation}
	S_{26} =
	\left(
	\begin{array}{ccccc}
		0 \; & 1 \; & 22 \; & 22 \; & 9 \\
		1 \; & 0 \; & 22 \; & 22 \; & 9 \\
		22 \; & 22 \; & 0 \; & 5 \; & 22 \\ 
		22 \; & 22 \; & 5 \; & 0 \; & 22 \\
		9 \; & 9 \; & 22 \; & 22 \; & 0
	\end{array}
	\right) ;\qquad \qquad
	E_{26} =
	\left(
	\begin{array}{c}
		0 \\
		0 \\
		0 \\
		0 \\
		20
	\end{array}
	\right).
	\label{mat26e}
\end{equation}

One important case occurs when two species merge into a single species (speciation reversal). This
might happen, for instance, when two species that have just become reproductively isolated are able to
breed again because of a mutation. The resulting merged species will have individuals with parents in
both ancestral species and we need to define which one 'survived' and which went extinct. Although
this is just a matter of labeling the species, we call the surviving species the one with most parents in
the previous generation. 

The drawing of species phylogenies for SSEE matrices is almost identical to that for MRCAT matrices.
The only differences are that nodes represent species, not individuals, and branches associated to
extinct species should not be drawn all the way down to present time, but should stop at the
extinction time recorded in the vector E. As in the MRCAT case of separation of lineages by sex, a
unique tree is generated independently of the clustering procedure chosen, due to the exact times of
speciation and extinction recorded in simulations based on this algorithm.

\subsection{Phylogenies generated by ancestor-descendant relationships (MRCAT) versus trees from speciation and extinction events (SSEE)}
\label{tree}

At the end of a simulation the MRCAT matrix contains the exact time to the most recent common
ancestor between every pair of individuals in the population. The SSEE matrix contains the
equivalent information at the species level, including extinct species. Both these matrices can be
used to draw phylogenetic trees. To draw a phylogeny of species considering the ancestral-descendant
relationships between individuals we can use the MRCAT matrix with the following reasoning: if $N_S$
species exist at time $t$ and $ind(i,j)$ is the $j$-th individual of the $i$-th species, a $N_S
\times N_S$ sub-matrix of the full MRCAT matrix can be generated considering only one individual per
species (Fig. \ref{figMRCAT}); a simple choice is to take $ind(i,1)$ for $i=1, 2, \dots N_S$ so that $T^{phy}_{i,j}
\equiv T_{ind(1,i),ind(1,j)}$.
 
\begin{figure}
	\centering \includegraphics[scale=0.5]{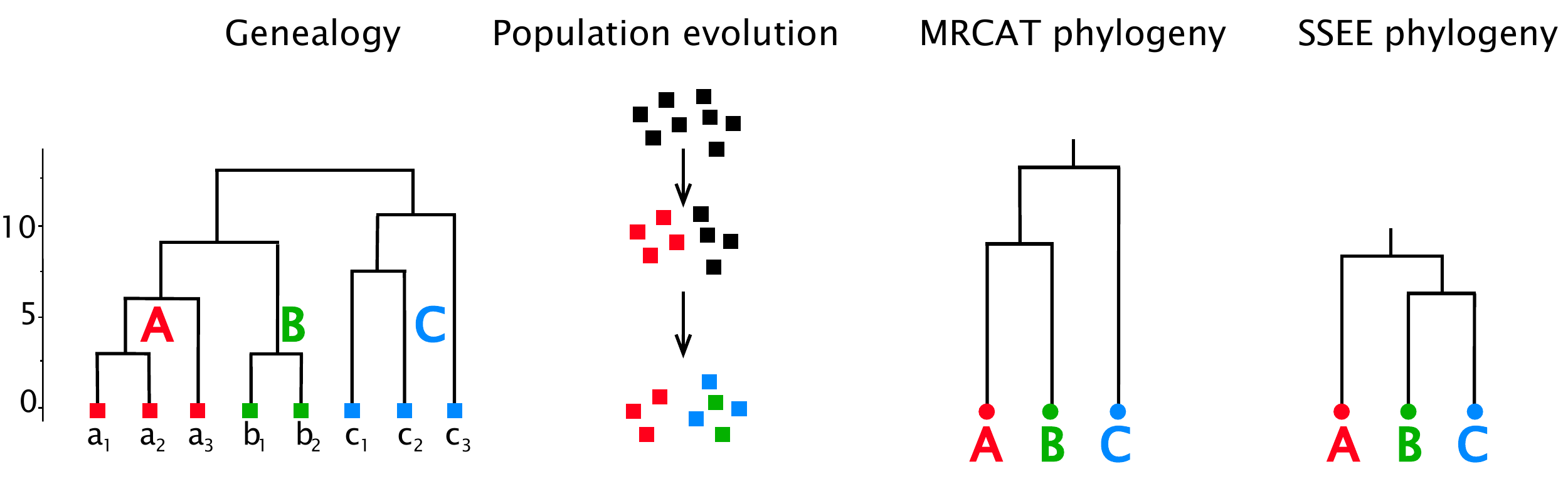} 
	\caption{Illustration of a genealogy recorded with MRCAT and the corresponding population evolution.
		The phylogenies constructed via MRCAT and SSEE differ in this case because, although individuals from
		species A and B have a more recent common ancestor than with individuals in C, species A split first,
		followed by the separation of B and C.} 
	\label{figcomp}
\end{figure}

The tree drawn from the SSEE algorithm is the true phylogeny of species, because it record the exact
speciation and extinction events, representing the actual branching process. On the other hand, the
phylogeny of species drawn from the MRCAT algorithm is different, although similar, from the true
phylogeny, because the time to the most recent common ancestor between individuals of different
species is only an approximation to the speciation time, since speciation can happen several
generations later. Figure \ref{figcomp} illustrates this situation: if a population splits into
three species in two closely spaced speciation events, it might happen that the first group to
speciate, species A in the figure, has a more recent common ancestor with the subgroup B than B with
C. During the time when B and C still form a single species reproduction between their individuals
might not happen for a while until they split, preserving the long time ancestry. This is more
likely to happen in populations with a spatial structure when individuals belonging to the two
subpopulations occupy different areas.

%

\section{Applications of MRCAT and SSEE algorithms to an individual-based model}
\label{app}

\subsection{The speciation model}
\label{example}

The model considered here to exemplify the MRCAT and SSEE algorithms is an extension of the
speciation model introduced in \cite{de2009global} and adapted in \cite{baptestini2013role} to
characterize individuals with separated sexes (males and females). The model has already been
studied in terms of speciation rates, species-area relationships and species abundance
distributions. Here we are adding the historical information generated by MRCAT and SSEE algorithms,
i.e., recording the parenthood of individuals from one generation to another (genealogy) as well as
the pattern and time of the speciation and extinction events (phylogeny or time tree).

The model describes a population of $N$ haploid individuals that are genetically identical at the
beginning of the simulation and are randomly distributed in a $L \times L$ spatial lattice with
periodic boundary conditions. More than one individual is allowed in each site of the lattice, but
because the density of the population is low, this seldom occurs. The genome of each
individual is represented by a sequence of $B$ binary {\it loci}, with state 0 or 1, where each {\it
	locus} plays the role of an independent biallelic gene. Individuals also carry one separate label
that specify their sex, male or female. The evolution of the population involves the combined
influence of sexual reproduction, mutation and dispersal \cite{de2009global}.

The reproduction trial starts with individual 1 and goes to individual $N$, so that all individuals
of the population have a chance to reproduce. The individual selected for reproduction, the {\it
	focal individual}, searches for potential mates in its {\it mating range}, a circular area of radius
$S$ centered on its spatial location. The focal individual can only reproduce with those within its
mating range and if they are genetically compatible, i.e., if the genetic distance between them is below
a particular threshold $G$. Among the compatible individuals within its mating range one of the
opposite sex is randomly chosen as mating partner. Individuals whose genetic distance is larger than
$G$ are considered reproductively isolated (threshold effect \cite{gavrilets2014models}). Genetic
distances between individuals are calculated as the Hamming distance \cite{hamming1950error} between
their genetic sequences, i.e., the number of loci at which the corresponding alleles are different.

Once the focal individual finds a compatible mate of the opposite sex, reproduction proceeds
with the combination of their genetic materials to produce the offspring genome, with each {\it
	locus} having an equal probability of being transmitted from mother or father. After combination
each {\it locus} in the offspring genome can mutate with probability $\mu$. Finally, the offspring
replaces the focal reproducing individual. In each reproductive event only one descendant is
generated. The offspring is then dispersed with probability $D$ within a region around the expiring
focal parent. There is a probability $Q$ that the focal individual will die without reproducing. In
this case a neighbor is randomly selected from its mating range to reproduce in its place, so that
the population size remains constant.

Evolution proceeds in non-overlapping discrete generations such that the entire population is
replaced by offspring. Species are defined as groups of individuals connected by gene flow, so that
any pair of individuals belonging to different species are reproductively isolated (genetic distance
greater than $G$). However, two individuals belonging to the same species can also be reproductively
isolated, as long as they can exchange genes indirectly through other individuals of the species.


\subsection{Phylogenies estimated based on genetic distances}
\label{gdm}

As we have described in the previous subsection, the genome of all individuals are identical at the
beginning of the simulation but mutations introduce differences and after many generations the
population will display a distribution of genomes. Genetic distances can, therefore, be calculated
between pairs of individuals and be used as a proxy for ancestry, such that the larger the genetic
distance between two individuals the farther back should their common ancestor. In order to estimate
phylogenies by genetic distance, we selected the same individuals per species that was used in the
drawing of the phylogeny via MRCAT and computed a matrix of genetic distances. This process mimics
the sampling of individuals from a real population and the comparison of their DNA's as a measure of
ancestry.

From the genetic distance matrix, we estimated trees from three distance-based methods. Firstly, we
used a hierarchical clustering method to cluster the species based on the genetic distances. The
clustering was realized first using the algorithm UPGMA \cite{murtagh1984complexities}. In this
algorithm two groups of species are clustered based on the average distance between all members of
the groups. This method assumes a constant rate of change, generating ultrametric trees in which
distances from the root to all tips are equal. Secondly, we used the NJ method
\cite{saitou1987neighbor} of phylogenetic inference. In this method the procedure is to find pairs
of neighbors in which the total branch length at each stage of the clustering is minimal, starting
with a starlike tree. Finally, we used the ME method \cite{rzhetsky1993theoretical}, which assumes
that the true phylogeny is probably the one with the smallest sum of branch lengths, as in the NJ
method. The difference is that in the ME method a NJ tree is constructed first and next tree
topologies close to this NJ tree are estimated by certain criteria, with all these trees being
examined and the tree with the small sum of branch lengths being chosen. We used the function $hclust$ of
the stats package \cite{rlanguage} to estimate ultrametric trees from the UPGMA method. To estimate trees from the NJ
method, we used the $nj$ function of the ape package \cite{paradis2004ape}. In this case, the
estimated trees are not ultrametric, so we transform then in ultrametric trees using the $chronoMPL$ and $multi2di$ functions
in ape package \cite{paradis2004ape,britton2002phylogenetic}. We used the $Rkitsch$ function of the Rphylip package
\cite{revell2014rphylip,felsenstein2002phylip} to estimate ultrametric trees from the ME method
assuming an evolutionary clock. The NJ and ME methods generally is considered superior to UPGMA
because it optimizes a tree according to minimum evolution criterion. Similarly to the UPGMA, the NJ
and ME methods are fast and efficient computationally.


\subsection{Statistical indexes to compare phylogenies}
\label{stats}

To evaluate the accuracy of the phylogenies generated by the MRCAT algorithm and by the genetic
distance methods (UPGMA, FM and ME) in relation to the true phylogeny generated by SSEE we use three
statistics: the Robinson and Foulds (RF \cite{robinson1981comparison}) metric, the gamma statistic
($\gamma$ \cite{pybus2000testing}) and the Sackin's index ($I_s$ \cite{sackin1972good,blum2005statistical}).

The RF metric measures the distance between phylogenetic trees, providing the overall topological resemblance of the phylogenies. Specifically, the RF metric calculates the number of internal branches present in only one of the trees being compared. If we consider two trees, $T1$ and $T2$, we have:
\begin{equation}
RF(T_1,T_2) = \frac{L_1}{L'_1} + \frac{L_2}{L'_2}
\label{rf}
\end{equation}
in which $L_1$ and $L_2$ are the number of branches on $T_1$ and $T_2$, respectively. The number of branches shared by $T_1$ and $T_2$ are represented by $L'_1$ and $L'_2$.
The RF metric was calculated using the $RF.dis$ function of the phangorn package \cite{schliep2011phangorn}.

The $\gamma$-statistic measures the distribution of branch lengths of a tree and is defined as \cite{pybus2000testing}:
\begin{equation}
\gamma = \frac{1}{D} \left[ \frac{1}{N_S-2} \sum_{k=2}^{N_S-1} T(k) - T(N_S)/2 \right]
\label{gamma}
\end{equation}
where
\begin{equation}
T(k) = \sum_{j=2}^{i} j g_j;
\end{equation}
\begin{equation}
D = T(N_S) / \sqrt{12(N_S-2)}
\end{equation}
and $g_k$ is the time interval between speciation events as represented by the nodes of the tree.
Under a pure birth process with constant speciation rate per branch, $\gamma$-values follow a
standard normal distribution centered on $\gamma=0$ with unit standard deviation. If $\gamma>0$ the
internal nodes are closer to the tips and if $\gamma<0$ they are closer to the root, as compared
with the case of constant speciation rate. The $\gamma$-statistic was calculated using the 
$gammaStat$ function of the ape package \cite{paradis2004ape}.

The Sackin's index measures the degree of imbalance, or asymmetry, of a tree
\cite{sackin1972good,blum2005statistical}. It is defined as 
\begin{equation}
I_s = \sum_{j} d_j 
\label{sackin}
\end{equation}
in which $d_j$ is the number of nodes to be traversed between each leaf $j$ and the root, including
the root \cite{dearlove2015measuring}. The expected Sackin's index under a pure birth process (the Yule model \cite{yule1925mathematical}) is 
\begin{equation}
 E(I_s(N_S)) = 2N\sum_{k=2}^{N_S}\frac{1}{k}
\end{equation}
where $N_S$ is the number of leaves. The expected Sackin's index approximates $2N_S\log{N_S}$ for large
values of $N_S$ \cite{blum2005statistical}. Since the expected value of the Sackin's index increases
with the tree size, a normalized index is defined to compare trees with different sizes: 
\begin{equation}
I^n_s = \frac{I_s(N_S) - E(I_s(N_S))}{N_S}.
\end{equation}
Here we used the normalized Sackin's index to compare the phylogenies. The Sackin's index was calculated using the 
$sackin$ function of the apTreeshape package \cite{aptreeshape}.


\section{Results}
\label{results}

\begin{figure}
	\centering \includegraphics[scale=0.5]{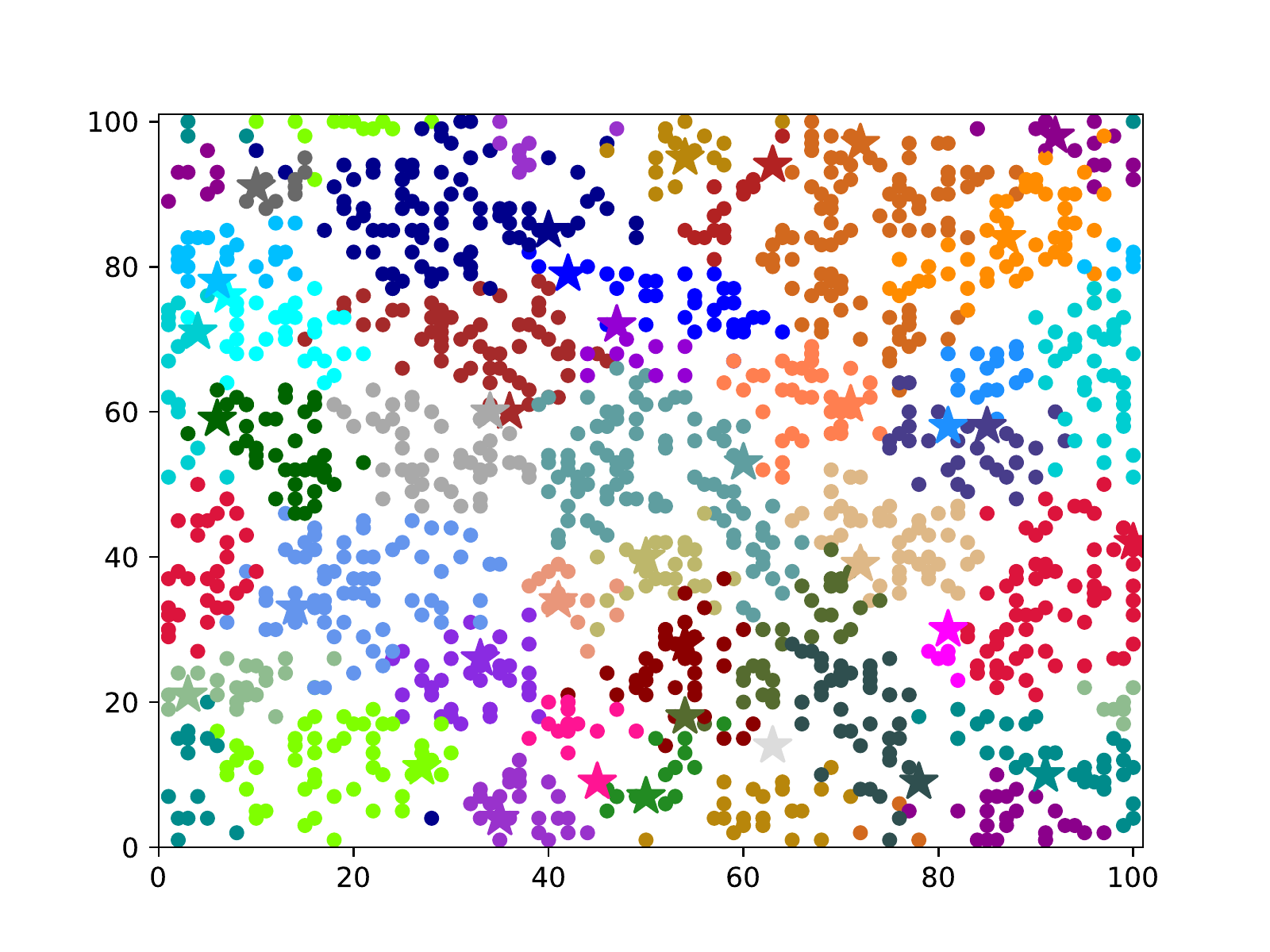} 
	\caption{Spatial distribution of individuals from one simulation based on the model described in section 
		\ref{example}. Individuals are represented by circles, and each color represents a different species.
		Stars indicate the individuals used to draw the phylogenies shown in figure \ref{fig6}.} 
	\label{fig4}
\end{figure}

We have ran simulations of the speciation model described in section \ref{example} with parameters
$N=1500$, $L=100$, $B=150$, $S=5$, $G=7$, $\mu=0.001$, $Q=0.05$. We start showing the results of a
single simulation to show examples of phylogenies. Figure \ref{fig4} shows the population after 1000
generations, with circles representing individuals and colors indicating the 36 species generated.
Species form spatial clusters, a consequence of the small $S$ value used the simulation.

\begin{figure}
	\centering{
		\includegraphics[scale=0.34]{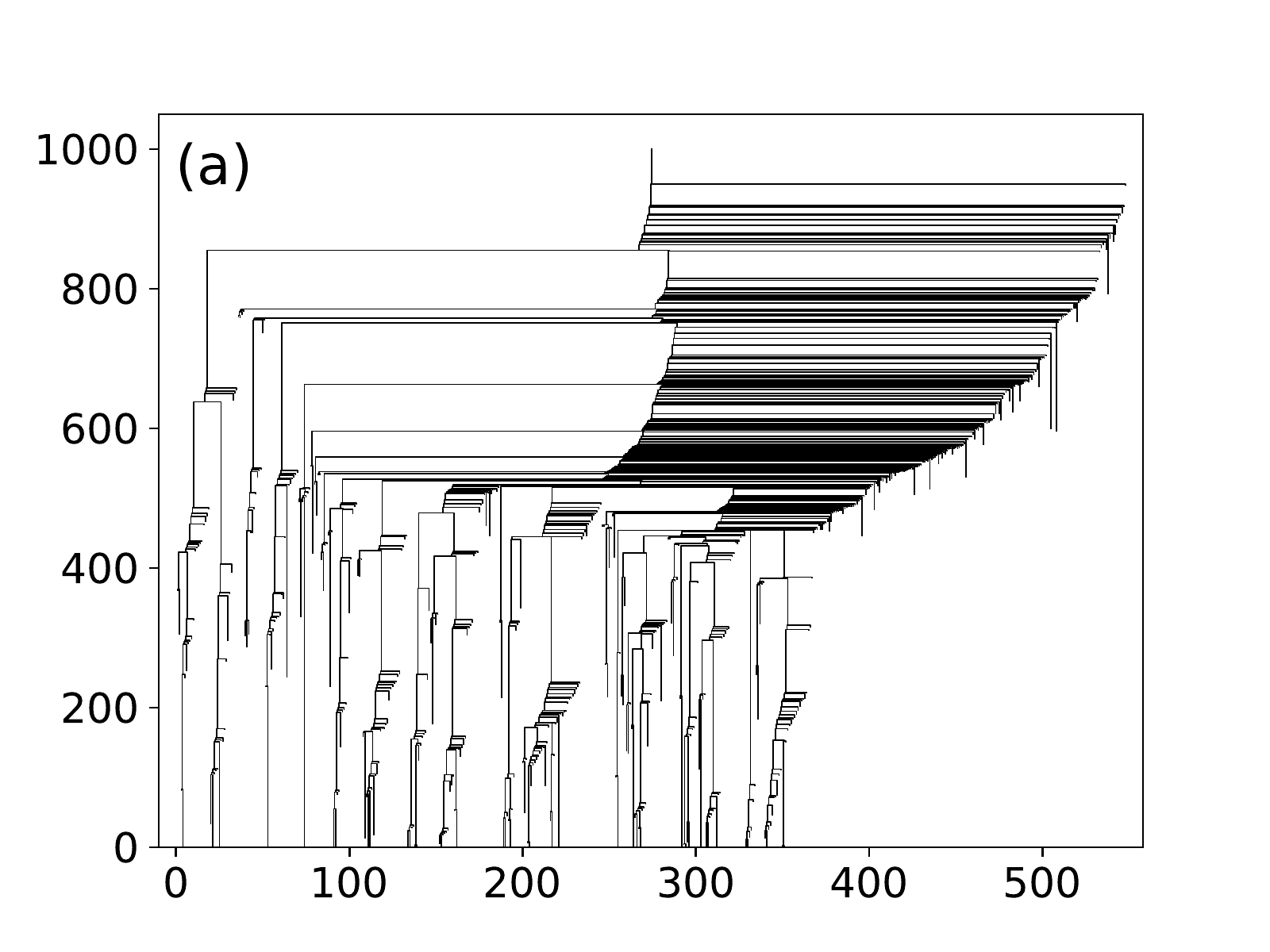} \qquad
		\includegraphics[scale=0.34]{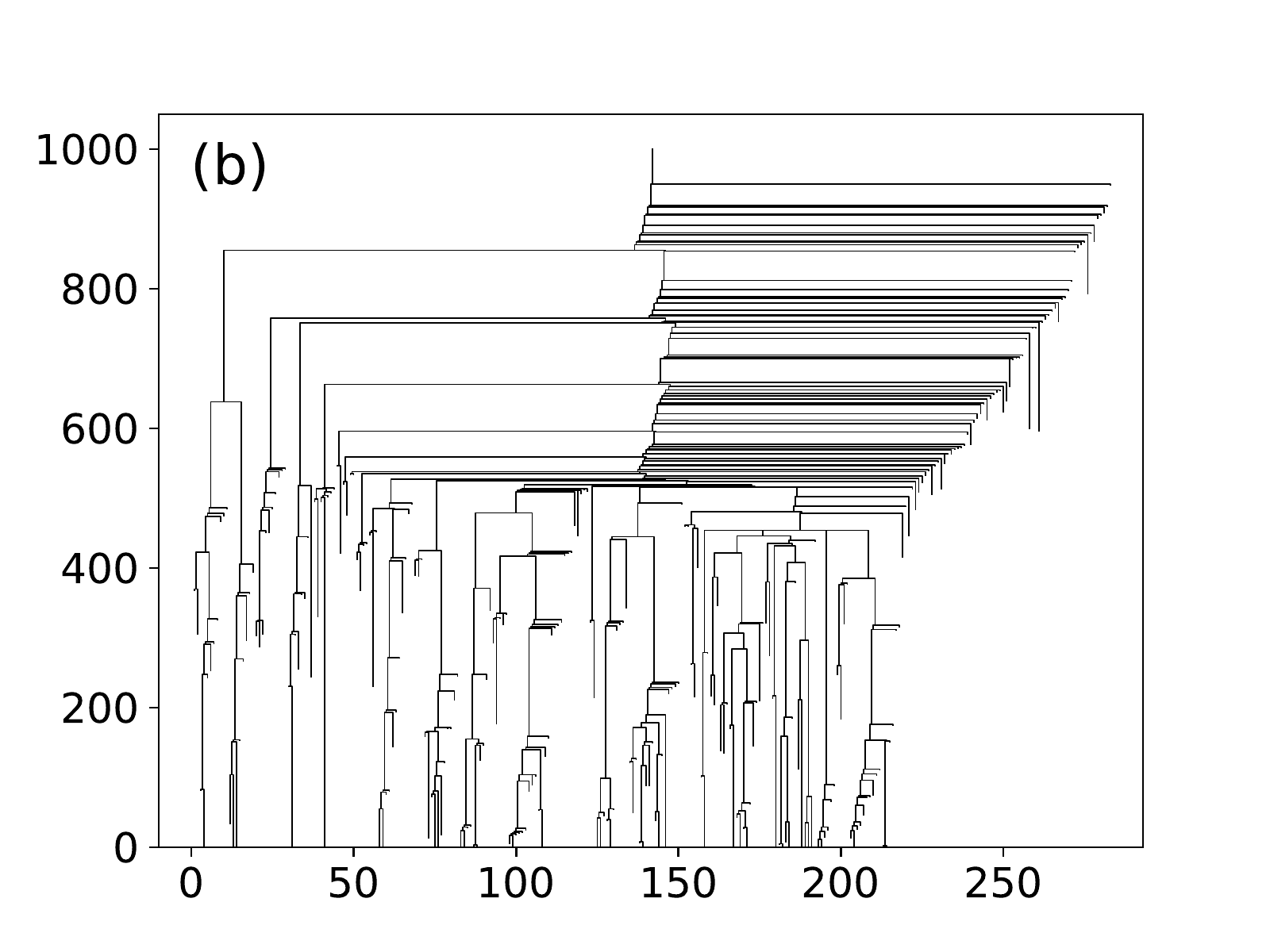} \vspace{0.5cm}
		\includegraphics[scale=0.34]{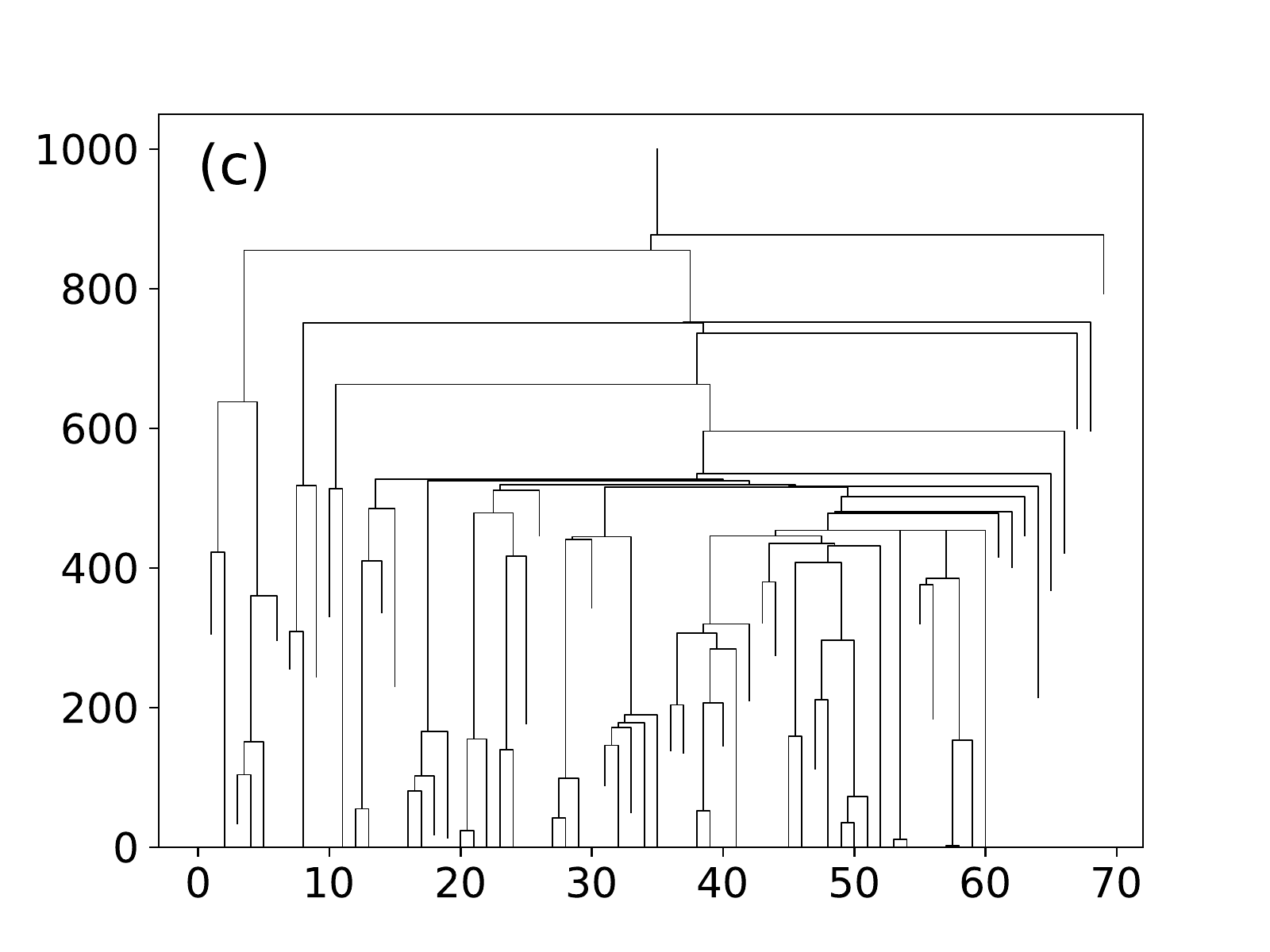} \qquad
		\includegraphics[scale=0.34]{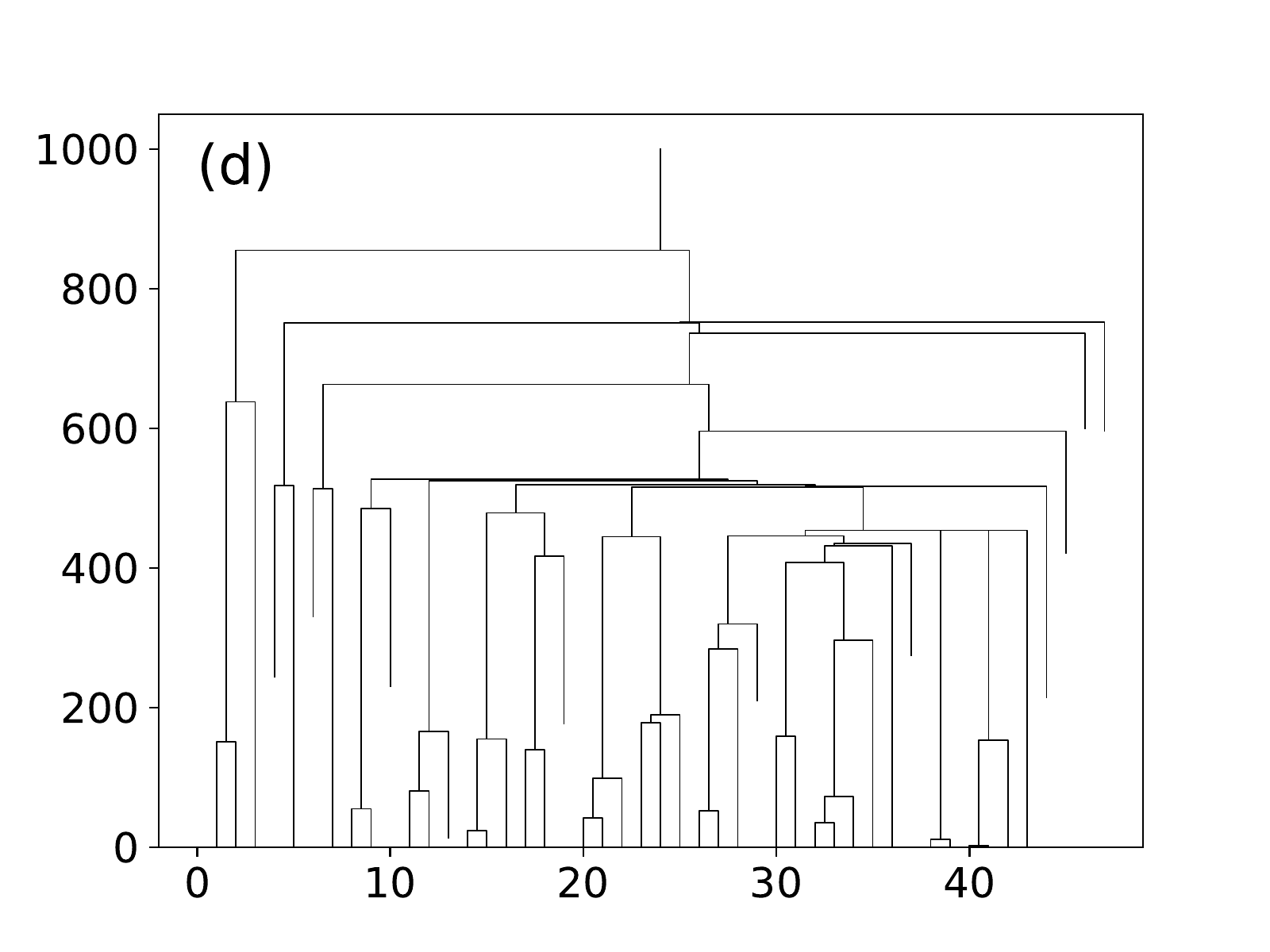} }
	\caption{True phylogenies obtained with the SSEE method. (a) full phylogeny, including all speciation and extinction events; (b) filtered phylogeny, excluding branches (species) which had more than 20 individuals at the moment of extinction; (c) filtered phylogeny, excluding also branches that lasted less than 50 generations and (d) 100 generations.} 
	\label{fig5}
\end{figure}

The true phylogenetic tree of the population, generated using the SSEE algorithm, is shown in Fig.
\ref{fig5}. Figure \ref{fig5}(a) shows the full phylogeny, which includes all speciation and
extinction events. The large number of events seen near the root of the tree correspond mostly to
unsuccessful or incomplete speciation events, in which a group of individuals momentarily splits in
two species but quickly recombine into a single species due to mutations. This phenomenon is very
common at the beginning of the speciation process in the model described in section \ref{example}.
In Fig. \ref{fig5}(b),(c),(d) the full phylogeny was filtered in order to remove speciation
reversals and keep only {\it true} extinction events. In the model, extinctions occur by stochastic
fluctuations in the number of individuals of a species, which might become very small and go to
zero. Figure \ref{fig5}(b) shows the phylogeny filtered by the criterion of population size at the
moment of vanishing: species that disappear with more than 20 individuals were considered
speciation reversals and removed from the tree. Figs. \ref{fig5}(c) and (d) display the same
phylogenies but filtered also by the criterion of persistence time: branches of species that lasted
less than 50 generations (c) or 100 generations (d) were removed.

\begin{figure}
	\centering{
		\includegraphics[scale=0.5]{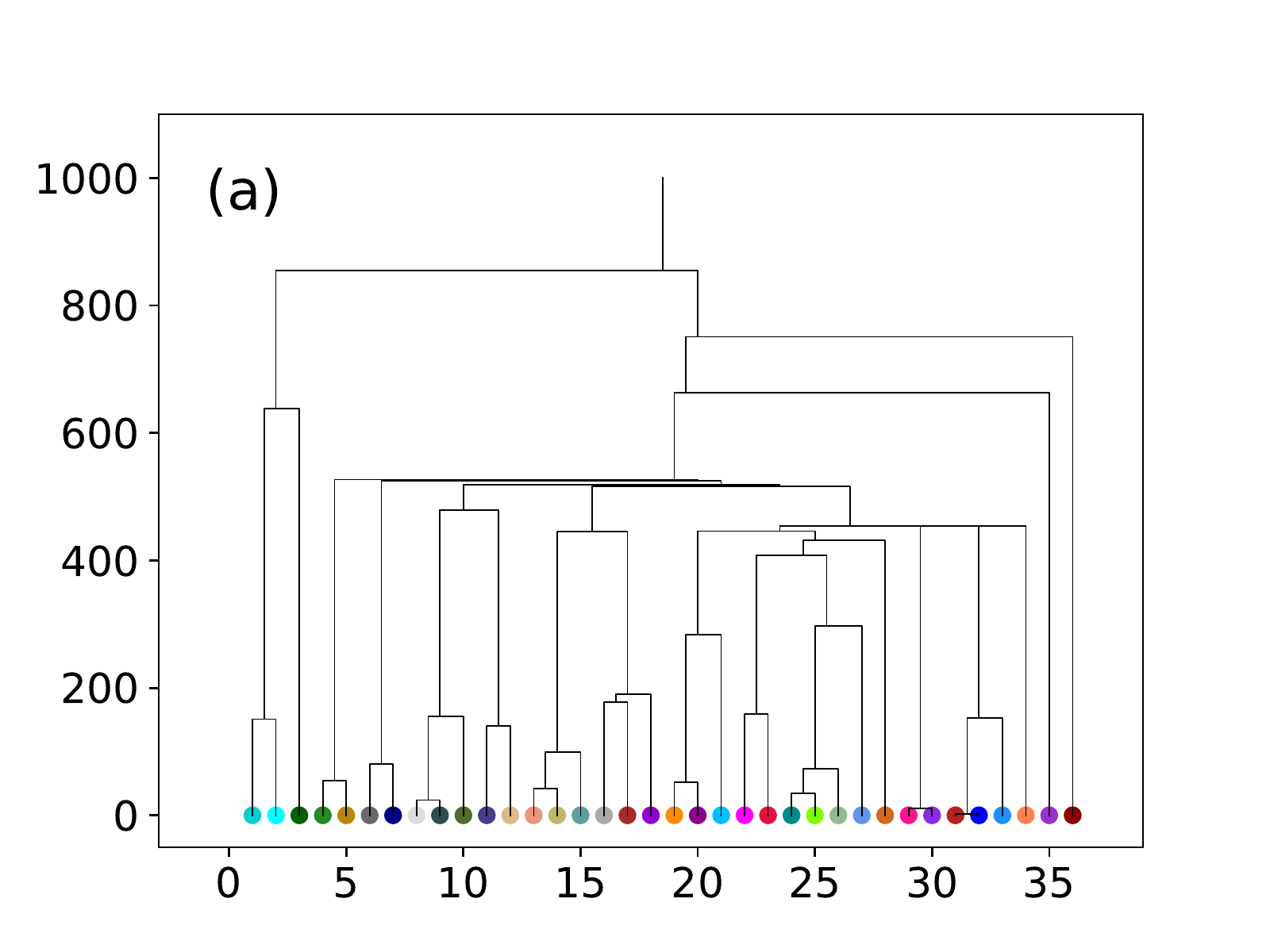} 
		\includegraphics[scale=0.5]{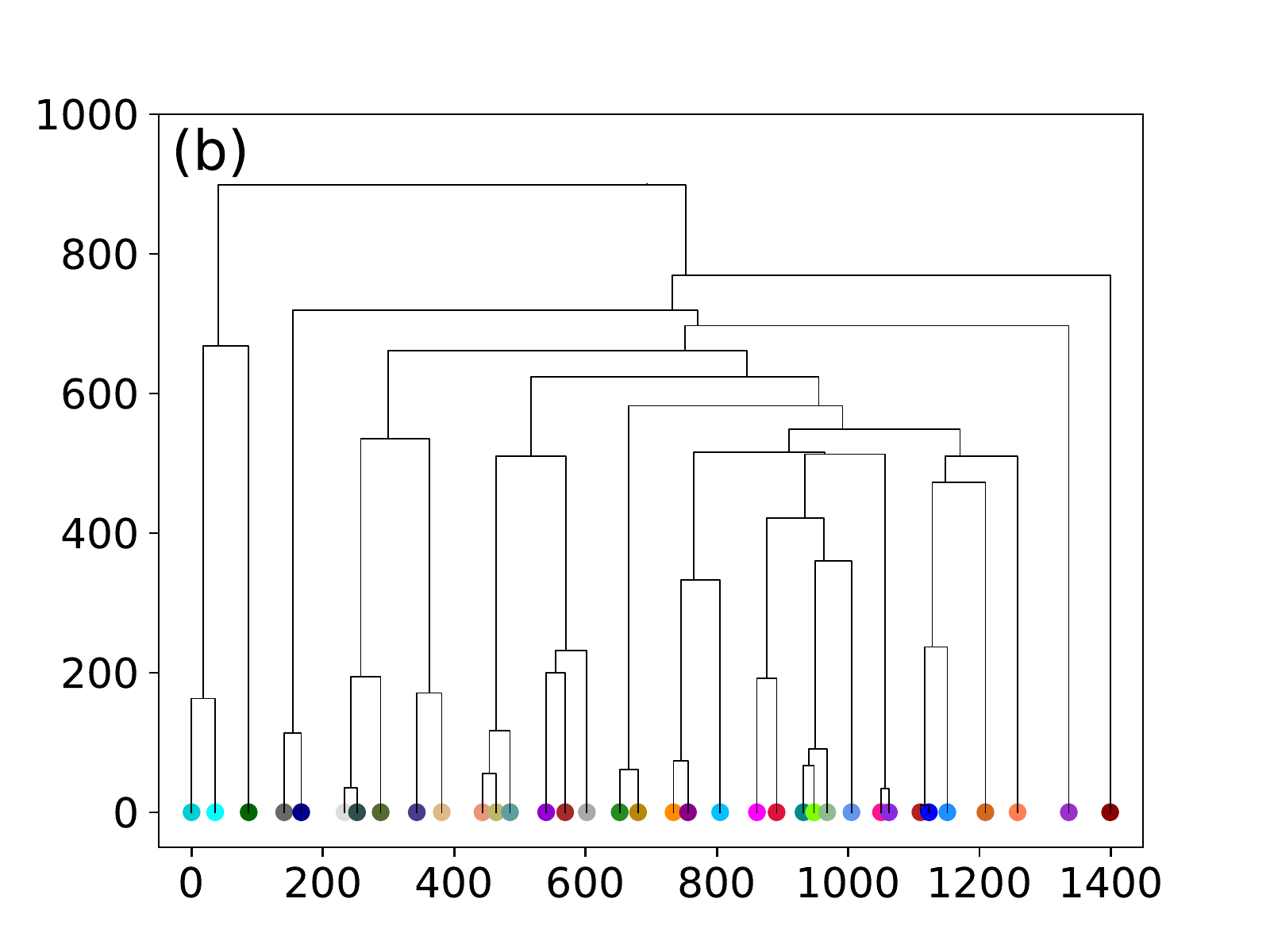} 
		\includegraphics[scale=0.5]{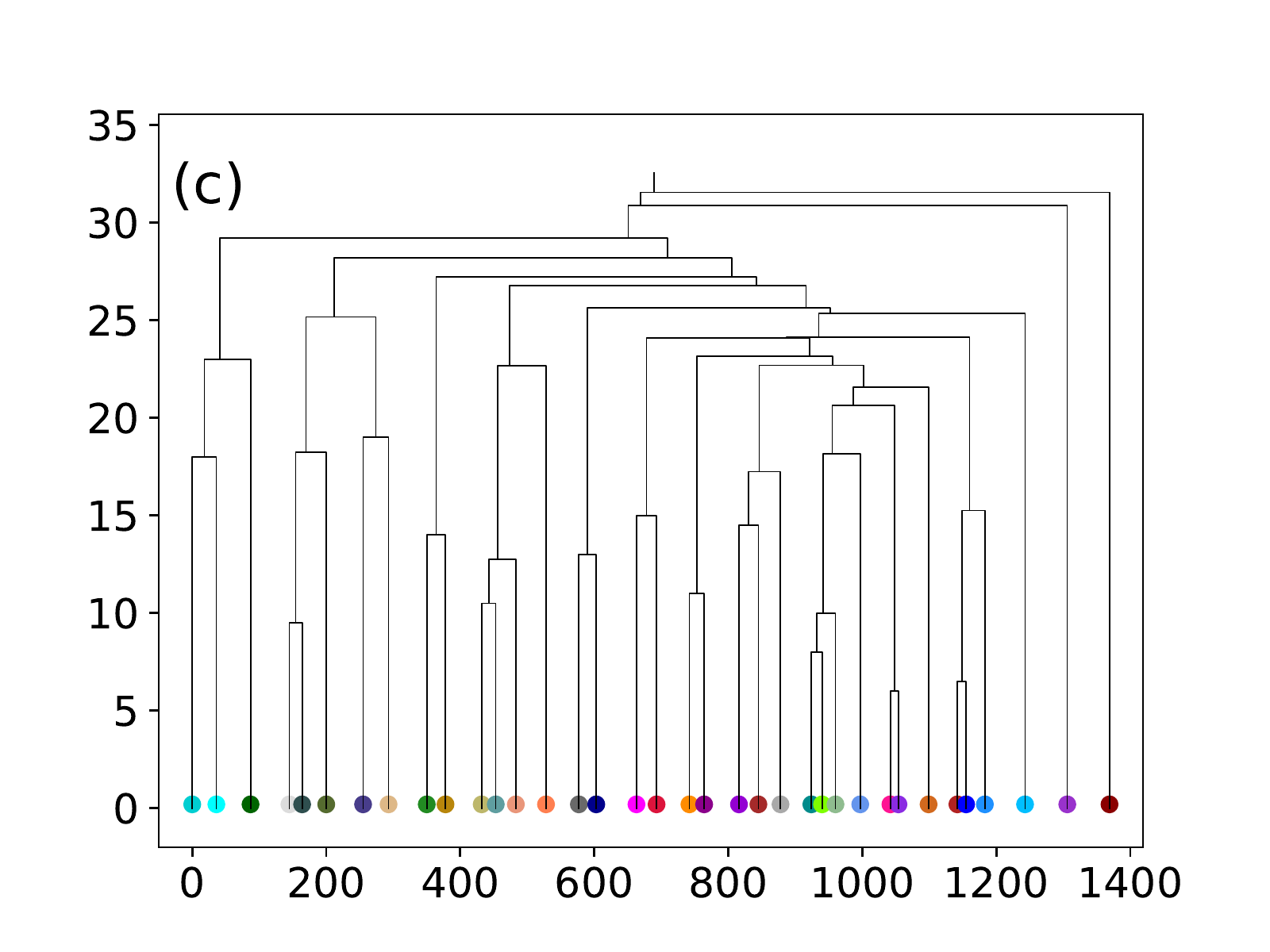}}
	\caption{(a) Extant phylogeny obtained via SSEE (species are separated by one unit on x-axis); (b) via MRCAT; (c) 
		via genetic distance matrix using UPGMA (neighbor species are separated by genetic distances). Colors correspond to 
		species in Fig. \ref{fig4}.} 
	\label{fig6}
\end{figure}

Phylogenies computed from the SSEE, MRCAT and genetic distance matrices are shown in Fig. \ref{fig6}. 
Panel (a) shows the true SSEE phylogeny, filtered to exhibit only the extant species. Panel (b) was 
obtained from the MRCAT algorithm, with one individual from each species 
being selected to represent the species. Finally panel (c) shows the phylogeny estimated from the 
genetic distance matrix of the same individuals used in Fig. \ref{fig6}(b) by the UPGMA clustering 
method. Differences in topology and branch lengths are qualitatively visible among these trees. 
Maternal and paternal genealogies obtained from the MRCAT algorithm are shown in Fig. S2.

\begin{figure}[!htpb]
	\centering{ \includegraphics[scale=0.48]{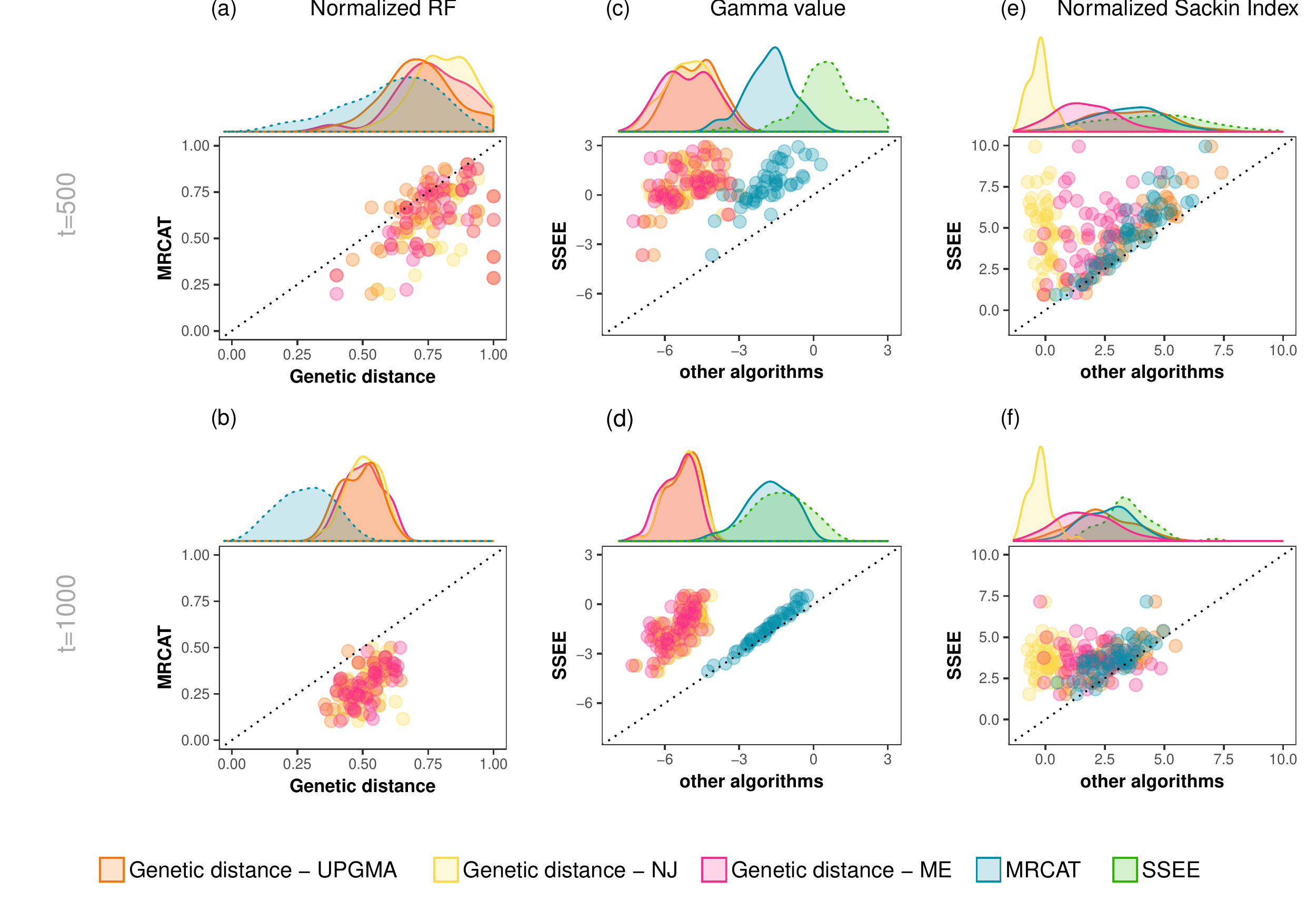} } 
	\caption{Comparisons among phylogenies generated by
	the algorithms proposed here (MRCAT and SSEE) and phylogenies estimated from genetic distance by
	UPGMA, NJ and ME methods. Lines exhibit the comparisons of RF, gamma and Sackin's metrics of 50
	simulations at times $500$ (first line) and $1000$ (second line) generations. Colors represent the
	different methods utilized to generate the trees. (a) and (b): difference between RF-values of
	phylogenies obtained by MRCAT ($y$-axis) and by genetic distance-based methods ($x$-axis). Small
	values of RF indicate that phylogenies are closer to the true phylogeny (SSEE). (c) and (d):
	difference between branch length distributions ($\gamma$) of phylogenies generated by SSEE
	($y$-axis, green distribution) and MRCAT algorithm (blue) or genetic distance-based methods
	(orange, yellow and pink) ($x$-axis). (e) and (f): the same as (c) and (d), but considering now the
	degree of imbalance (Sackin index). Distributions above all scatterplots illustrate
	qualitatively the differences in topology (a,b), branch length distribution (c,d) and degree of
	imbalance (e,f) of phylogenies generated from each algorithm or method in the 50 simulations.} 
\label{fig7}
\end{figure}

Statistical comparisons among phylogenies generated by the MRCAT algorithm and by the genetic
distance methods (UPGMA, NJ and ME) in relation to the true phylogeny (SSEE) are shown in Fig.
\ref{fig7}. The first line shows comparisons of topology (RF metric), branch length distribution
($\gamma$-statistic) and degree of imbalance (Sackin index) among phylogenies after 500 generations
in 50 simulations. The second line shows the same comparisons after 1000 generations for the same 50
simulations. Colors represent the different methods utilized to generate the trees. In the RF
scatterplots (Fig. \ref{fig7}(a)(b)) the coordinates of each point refer to the normalized
topological distance between the tree calculated with the MRCAT matrix ($y$-axis) or by genetic
distance matrix ($x$-axis) from the true phylogenies generated by the SSEE algorithm. Small values
of RF indicate that phylogenies are closer to the true phylogeny (SSEE). The diagonal dotted line
defines the condition in which the topology of the phylogenies (RF-value) was equal in trees
generated by genealogical relationships (MRCAT trees) and that estimated by genetic distance (
UPGMA, NJ and ME methods). The scatterplot for $T=500$ (Fig. \ref{fig7}(a)) shows that phylogenies
generated by MRCAT and genetic distance using UPGMA method (orange points) were similar in their
RF-values, while trees estimated from NJ and ME methods (yellow and pink) had more different
RF-values. For $T=1000$ (Fig. \ref{fig7}(b)) all phylogenies estimated by genetic distance-based
methods differ from those obtained by MRCAT. The density distribution of RF values shown above
the scatterplots indicate that MRCAT is always closer to SSEE, especially for $T=1000$.

Regarding the branch length distribution the scatterplots (Fig. \ref{fig7}(c),(d)) show the
difference between $\gamma$-values in SSEE phylogenies ($y$-axis) and MRCAT or genetic distance
(UPGMA, NJ or ME) phylogenies ($x$-axis). The diagonal dotted line defines the condition in which
the $\gamma$-values of trees generated by genealogical relationships (MRCAT trees) or by genetic
distance (by UPGMA, NJ and ME methods) were equal to values of true phylogenies. We observe that for
both times (Fig. \ref{fig7}(c),(d)) MRCAT trees had $\gamma$ distributions closer to true
phylogenies (SSEE) than all genetic distance-based trees, with a large match in $T=1000$ . Finally,
the normalized Sackin's index is presented in Fig. (Fig. \ref{fig7}(e),(f)). The imbalance of MRCAT
phylogenies was closer to the true phylogenies in $T=500$ (Fig. \ref{fig7}(e)). On the other hand,
for $T=1000$ the imbalance was similar for MRCAT and all distance-based methods, except for the NJ. 
The NJ trees exhibited the most incorrect Sackin index (Fig. \ref{fig7}(e)(f)), possibly because NJ 
trees are not rooted, a necessary condition to compute this index. The rooting procedure chosen can 
be quite arbitrary, affecting the balance of the trees and consequently the Sackin index.
The distributions above all scatterplots show qualitatively the differences in topology (Fig.
\ref{fig7}(a),(b)), branch length distribution (Fig. \ref{fig7}(c),(d)) and degree of imbalance
(Fig. \ref{fig7}(e),(f)) of phylogenies generated from each algorithm or method in the 50
simulations performed in each time ($t=500$ or $t=1000$).


\section{Discussion}
\label{discussion}

Understanding all the mechanisms that promote speciation is still an open problem in evolutionary
biology \cite{gavrilets2014models,kirkpatrick2002speciation}. Even more challenging is to identify
which of these mechanisms were important in a particular case. A large number of mathematical and
computational models were developed in the past years to evaluate hypothesis for speciation
processes involving different forces (neutral
\cite{hoelzer2008isolation,desjardins2011likely,melian2012does,baptestini2013conditions}, sexual
selection \cite{van2009origin,uyeda2009drift,m2012sexual}, ecological selection
\cite{rettelbach2013three,nosil2012ecological}), including in some models the role of geography in
speciation (allopatric \cite{fierst2010genetic,gourbiere2010species,
	fraisse2014genetics,yamaguchi2013first}, parapatric \cite{gavrilets2000patterns,bank2012limits}, and
sympatric speciation
\cite{m2012sexual,rettelbach2013three,burger2006conditions,pennings2007analytically}). The results
of models, however, can only seldom be compared with real data
\cite{bolnick2007sympatric,gavrilets2009adaptive}. In these cases comparisons are often made in a
macroecological scale, including qualitative species abundance and spatial distributions,
species-area relationships and genetic or phenotypic distances
\cite{de2009global,martins2013evolution,May20141657,Gomes2012,GarciaMartin05072006}. Nevertheless,
very little attention has been given to the evolutionary history of individuals and species from
these models, neglecting the macroevolutionary scale underlying the speciation process
\cite{manceau2015phylogenies,rosindell2015unifying}.

In this paper we have described two procedures to register the history of individuals (MRCAT) or
species (SSEE) in individual-based models. With the ancestral-descendant relationships or speciation
events saved in MRCAT and SSEE matrices we drawn trees using a clustering algorithm. These trees
have properties demonstrated in section I of Supporting Information. In the MRCAT algorithm,
genealogies of individuals and phylogenies of species were obtained, whereas in the SSEE algorithm
only phylogenies of species can be extracted. As in the SSEE algorithm the speciation events are
precisely recorded we have the true phylogenetic tree, whereas in the MRCAT algorithm the relations
among species are recovered from genealogical relationships between individuals. The MRCAT algorithm
allows the construction of maternal, paternal and general lineages, the last being analogous to
cases with hermaphroditic individuals. We have applied these algorithms to a spatially explicit IBM
where individuals are separated into males and females and sexual reproduction is restricted by
genetic similarity and spatial proximity. We showed that maternal, paternal and general genealogies
from the MRCAT algorithm are different even if the same individuals are chosen to draw the trees
(Supporting Information, section II). Maternal and paternal genealogies (Fig. S2(a),(b)) are
different because they constructed drawn from different MRCAT matrices. In the first case, the MRCAT
matrix contains the time to the most recent common $female$ ancestor between each pair of
individuals, while in the second case the MRCAT matrix has the time to the most recent common $male$
ancestor between the same individuals, which leads to different times and genealogical
relationships. In addition, for the general genealogy - taking the most recent common ancestor among
females and males ($i.e.$, disregarding sex) - the resulting MRCAT matrix does not uniquely specify
the genealogy (Fig. S2(c)). Regarding the phylogenetic trees, we showed that they may be different
if obtained by the MRCAT or SSEE algorithms (Fig. \ref{fig6}(a),(b), Fig. \ref{fig7}). As discussed
subsection \ref{tree} this mismatch happens because the time to the most recent common ancestor
between individuals of different species is only an approximation to the speciation time, since
speciation can happen several generations later (Fig. \ref{figcomp}).

Structural properties of the phylogenies, such as the Sackin index and Gamma distribution, obtained
from SSEE and MRCAT trees were compared to values calculated in phylogenies estimated from the
genetic distance among individuals of extant species by distance-based methods (UPGMA, NJ and ME).
The aim of this comparison was to show that the validity of these methods commonly used in empirical
studies, where the complete past history is inaccessible, can be assessed with the help of models.
Differences in topology and branch length distribution measured by the RF metric and
$\gamma$-statistic, respectively, revealed that MRCAT trees were closer to the true phylogenies
(SSEE) than genetic distance-based trees. This result can be derived from the chance that back
mutations happened in the genome of individuals, erasing the information needed to uncover the real
history among species. This phenomenon is more likely to happen at long times and for small genome
size. Indeed, we observed that in 500 generations (Fig. \ref{fig7}(a)(c)) the phylogenies estimated
from genetic distance were more closer to the ones generated from MRCAT algorithm than in 1000
generations (Fig. \ref{fig7}(b)(c)), possibly because in the first case back mutations were less
likely. Another factor to explain the difference between genetic distance-based and true phylogenies
is the sampling of only one individual to estimate the trees in the first case. Although this
sampling could be a reason for this difference it is not the only important factor, as phylogenies
generated from MRCAT algorithm also used only one individual per species - and the same individual
used to compute genetic distance - to drawn the trees and still were more closed to the true
phylogeny (Fig. \ref{fig7}(a),(b),(c),(d)). The degree of imbalance showed a different picture, with
less differences between MRCAT trees and genetic distance trees. Still, MRCAT trees are closer to
the true phylogenies than the others. Trees estimated from genetic information in IBMs should be closer to
the true phylogenies for larger genome sizes, where the probability of back mutations is smaller.
Individual-based models with large or infinite genome sizes already available
\cite{de2017speciation,higgs1992genetic} would provide good tests for measuring the accuracy of
trees obtained by distance-based methods.


\section{Conclusions}
\label{conclusions}

The recent interest in the role of evolutionary history to explain the spatial patterns of abundance
and species diversity calls for the incorporation of phylogenetic trees in the speciation modeling
approach. Phylogenetic trees are essential tools to understand these patterns of diversity. They
reveal how species are related to each other and the times between speciation events. Moreover,
topological structure and branch length also contain clues about processes originating a particular
group of species. Previous works have already considered this problem for simpler models where each
mutation corresponds directly to a new species \cite{manceau2015phylogenies}. Our study provides the
first general attempt to extend individual-based models to incorporate the branching process using
the ancestral-descendant relationships between individuals and species. We believe this methodology
will help predict and classify the macroevolutionary branching process, as well as the corresponding
macroecological patterns ($e.g.$, species abundance distributions), resulting from different
speciation models. The comparison of these results with empirical studies may clarify the role of
different processes in generating the patterns observed in nature
\cite{turelli2001theory,field2009spatial}. Finally, the role of extinction in determining
macroevolutionary patterns is an open field \cite {quental2011molecular} which could be explored by
using the full phylogenetic trees generated from the SSEE algorithm.

\section*{Acknowledgments} 

This research was supported by Sao Paulo Research Foundation (FAPESP), National Council for Scientific and 
Technological Development (CNPq), and Coordination for the Improvement of Higher Education Personnel (CAPES). 
We thank P. L. Costa for his constructive comments about the manuscript, and A. B. Martins and L. D. Fernandes, who provided 
expertise that greatly assisted the research, all contributing with their insights and comments about the research results.

Conflicts of interest: none.

\bibliographystyle{elsarticle-num}

\begin{thebibliography}{10}
	\expandafter\ifx\csname url\endcsname\relax
	\def\url#1{\texttt{#1}}\fi
	\expandafter\ifx\csname urlprefix\endcsname\relax\def\urlprefix{URL }\fi
	\expandafter\ifx\csname href\endcsname\relax
	\def\href#1#2{#2} \def\path#1{#1}\fi
	
	\bibitem{coyne2004speciation}
	J.~A. Coyne, H.~A. Orr, Speciation. sunderland, ma (2004).
	
	\bibitem{de2009global}
	M.~A.~M. De~Aguiar, M.~Baranger, E.~Baptestini, L.~Kaufman, Y.~Bar-Yam, Global
	patterns of speciation and diversity, Nature 460~(7253) (2009) 384--387.
	
	\bibitem{gavrilets2014models}
	S.~Gavrilets, Models of speciation: Where are we now?, Journal of heredity
	105~(S1) (2014) 743--755.
	
	\bibitem{turelli2001theory}
	M.~Turelli, N.~H. Barton, J.~A. Coyne, Theory and speciation, Trends in Ecology
	\& Evolution 16~(7) (2001) 330--343.
	
	\bibitem{field2009spatial}
	R.~Field, B.~A. Hawkins, H.~V. Cornell, D.~J. Currie, J.~A.~F. Diniz-Filho,
	J.-F. Gu{\'e}gan, D.~M. Kaufman, J.~T. Kerr, G.~G. Mittelbach, T.~Oberdorff,
	et~al., Spatial species-richness gradients across scales: a meta-analysis,
	Journal of Biogeography 36~(1) (2009) 132--147.
	
	\bibitem{martins2013evolution}
	A.~B. Martins, M.~A. de~Aguiar, Y.~Bar-Yam, Evolution and stability of ring
	species, Proceedings of the National Academy of Sciences 110~(13) (2013)
	5080--5084.
	
	\bibitem{May20141657}
	F.~May, A.~Huth, T.~Wiegand, Moving beyond abundance distributions: neutral
	theory and spatial patterns in a tropical forest, Proceedings of the Royal
	Society of London B: Biological Sciences 282~(1802).
	
	\bibitem{Kopp2010}
	M.~Kopp, {Speciation and the neutral theory of biodiversity: Modes of
		speciation affect patterns of biodiversity in neutral communities.},
	BioEssays : news and reviews in molecular, cellular and developmental biology
	32~(7) (2010) 564--70.
	
	\bibitem{hubell2001unt}
	S.~P. Hubbell, The Unified Neutral Theory of Biodiversity and Biogeography,
	Princeton University Press, Princeton, NJ, 2001.
	
	\bibitem{gavrilets2000patterns}
	S.~Gavrilets, H.~Li, M.~D. Vose, Patterns of parapatric speciation, Evolution
	54~(4) (2000) 1126--1134.
	
	\bibitem{Dieckmann1999}
	U.~Dieckmann, M.~Doebeli, {On the origin of species by sympatric speciation.},
	Nature 400~(6742) (1999) 354--7.
	
	\bibitem{rettelbach2013three}
	A.~Rettelbach, M.~Kopp, U.~Dieckmann, J.~Hermisson, Three modes of adaptive
	speciation in spatially structured populations, The American Naturalist
	182~(6) (2013) E215--E234.
	
	\bibitem{gavrilets2003perspective}
	S.~Gavrilets, Perspective: models of speciation: what have we learned in 40
	years?, Evolution 57~(10) (2003) 2197--2215.
	
	\bibitem{Gomes2012}
	P.~R.~A. Campos, E.~D.~C. Neto, V.~M.~d. Oliveira, M.~A.~F. Gomes, Neutral
	communities in fragmented landscapes, Oikos 121~(11) (2012) 1737--1748.
	
	\bibitem{GarciaMartin05072006}
	H.~García~Martín, N.~Goldenfeld, On the origin and robustness of power-law
	species–area relationships in ecology, Proceedings of the National Academy
	of Sciences 103~(27) (2006) 10310--10315.
	\newblock \href
	{http://arxiv.org/abs/http://www.pnas.org/content/103/27/10310.full.pdf}
	{\path{arXiv:http://www.pnas.org/content/103/27/10310.full.pdf}}.
	
	\bibitem{manceau2015phylogenies}
	M.~Manceau, A.~Lambert, H.~Morlon, Phylogenies support out-of-equilibrium
	models of biodiversity, Ecology letters 18~(4) (2015) 347--356.
	
	\bibitem{pigot2010shape}
	A.~L. Pigot, A.~B. Phillimore, I.~P. Owens, C.~D.~L. Orme, The shape and
	temporal dynamics of phylogenetic trees arising from geographic speciation,
	Systematic Biology (2010) syq058.
	
	\bibitem{hagen2015age}
	O.~Hagen, K.~Hartmann, M.~Steel, T.~Stadler, Age-dependent speciation can
	explain the shape of empirical phylogenies, Systematic Biology 64~(3) (2015)
	432--440.
	
	\bibitem{quental2011molecular}
	T.~B. Quental, C.~R. Marshall, The molecular phylogenetic signature of clades
	in decline, PloS one 6~(10) (2011) e25780.
	
	\bibitem{davies2011neutral}
	T.~J. Davies, A.~P. Allen, L.~Borda-de {\'A}gua, J.~Regetz, C.~J. Meli{\'a}n,
	Neutral biodiversity theory can explain the imbalance of phylogenetic trees
	but not the tempo of their diversification, Evolution 65~(7) (2011)
	1841--1850.
	
	\bibitem{rosindell2015unifying}
	J.~Rosindell, L.~J. Harmon, R.~S. Etienne, Unifying ecology and macroevolution
	with individual-based theory, Ecology letters 18~(5) (2015) 472--482.
	
	\bibitem{deangelis2014individual}
	D.~L. DeAngelis, V.~Grimm, Individual-based models in ecology after four
	decades, F1000Prime Rep 6~(39) (2014) 6.
	
	\bibitem{murtagh1984complexities}
	F.~Murtagh, Complexities of hierarchic clustering algorithms: State of the art,
	Computational Statistics Quarterly 1~(2) (1984) 101--113.
	
	\bibitem{saitou1987neighbor}
	N.~Saitou, M.~Nei, The neighbor-joining method: a new method for reconstructing
	phylogenetic trees., Molecular biology and evolution 4~(4) (1987) 406--425.
	
	\bibitem{rzhetsky1993theoretical}
	A.~Rzhetsky, M.~Nei, Theoretical foundation of the minimum-evolution method of
	phylogenetic inference., Molecular biology and evolution 10~(5) (1993)
	1073--1095.
	
	\bibitem{higgs1992genetic}
	P.~G. Higgs, B.~Derrida, Genetic distance and species formation in evolving
	populations, Journal of molecular evolution 35~(5) (1992) 454--465.
	
	\bibitem{baptestini2013role}
	E.~M. Baptestini, M.~A. de~Aguiar, Y.~Bar-Yam, The role of sex separation in
	neutral speciation, Theoretical ecology 6~(2) (2013) 213--223.
	
	\bibitem{hamming1950error}
	R.~W. Hamming, Error detecting and error correcting codes, Bell Labs Technical
	Journal 29~(2) (1950) 147--160.
	
	\bibitem{rlanguage}
	{R Core Team}, \href{https://www.R-project.org/}{R: A Language and Environment
		for Statistical Computing}, R Foundation for Statistical Computing, Vienna,
	Austria (2017).
	\newline\urlprefix\url{https://www.R-project.org/}
	
	\bibitem{paradis2004ape}
	E.~Paradis, J.~Claude, K.~Strimmer, Ape: analyses of phylogenetics and
	evolution in r language, Bioinformatics 20~(2) (2004) 289--290.
	
	\bibitem{britton2002phylogenetic}
	T.~Britton, B.~Oxelman, A.~Vinnersten, K.~Bremer, Phylogenetic dating with
	confidence intervals using mean path lengths, Molecular phylogenetics and
	evolution 24~(1) (2002) 58--65.
	
	\bibitem{revell2014rphylip}
	L.~J. Revell, S.~A. Chamberlain, Rphylip: an r interface for phylip, Methods in
	Ecology and Evolution 5~(9) (2014) 976--981.
	
	\bibitem{felsenstein2002phylip}
	J.~Felsenstein, $\{$PHYLIP$\}$(phylogeny inference package) version 3.6 a3.
	
	\bibitem{robinson1981comparison}
	D.~F. Robinson, L.~R. Foulds, Comparison of phylogenetic trees, Mathematical
	biosciences 53~(1-2) (1981) 131--147.
	
	\bibitem{pybus2000testing}
	O.~G. Pybus, P.~H. Harvey, Testing macro--evolutionary models using incomplete
	molecular phylogenies, Proceedings of the Royal Society of London B:
	Biological Sciences 267~(1459) (2000) 2267--2272.
	
	\bibitem{sackin1972good}
	M.~Sackin, “good” and “bad” phenograms, Systematic Biology 21~(2)
	(1972) 225--226.
	
	\bibitem{blum2005statistical}
	M.~G. Blum, O.~Fran{\c{c}}ois, On statistical tests of phylogenetic tree
	imbalance: the sackin and other indices revisited, Mathematical biosciences
	195~(2) (2005) 141--153.
	
	\bibitem{schliep2011phangorn}
	K.~P. Schliep, phangorn: phylogenetic analysis in r, Bioinformatics 27~(4)
	(2011) 592--593.
	
	\bibitem{dearlove2015measuring}
	B.~L. Dearlove, S.~D. Frost, Measuring asymmetry in time-stamped phylogenies,
	PLoS computational biology 11~(7) (2015) e1004312.
	
	\bibitem{yule1925mathematical}
	G.~U. Yule, A mathematical theory of evolution, based on the conclusions of dr.
	jc willis, frs, Philosophical transactions of the Royal Society of London.
	Series B, containing papers of a biological character 213 (1925) 21--87.
	
	\bibitem{aptreeshape}
	N.~Bortolussi, E.~Durand, M.~Blum, O.~Francois,
	\href{https://CRAN.R-project.org/package=apTreeshape}{apTreeshape: Analyses
		of Phylogenetic Treeshape}, r package version 1.4-5 (2012).
	\newline\urlprefix\url{https://CRAN.R-project.org/package=apTreeshape}
	
	\bibitem{kirkpatrick2002speciation}
	M.~Kirkpatrick, V.~Ravign{\'e}, Speciation by natural and sexual selection:
	models and experiments, The American Naturalist 159~(S3) (2002) S22--S35.
	
	\bibitem{hoelzer2008isolation}
	G.~A. Hoelzer, R.~Drewes, J.~Meier, R.~Doursat, Isolation-by-distance and
	outbreeding depression are sufficient to drive parapatric speciation in the
	absence of environmental influences, PLoS Comput Biol 4~(7) (2008) e1000126.
	
	\bibitem{desjardins2011likely}
	P.~Desjardins-Proulx, D.~Gravel, How likely is speciation in neutral ecology?,
	The American Naturalist 179~(1) (2011) 137--144.
	
	\bibitem{melian2012does}
	C.~J. Meli{\'a}n, D.~Alonso, S.~Allesina, R.~S. Condit, R.~S. Etienne, Does sex
	speed up evolutionary rate and increase biodiversity?, PLoS Comput Biol 8~(3)
	(2012) e1002414.
	
	\bibitem{baptestini2013conditions}
	E.~M. Baptestini, M.~A. de~Aguiar, Y.~Bar-Yam, Conditions for neutral
	speciation via isolation by distance, Journal of theoretical biology 335
	(2013) 51--56.
	
	\bibitem{van2009origin}
	G.~S. van Doorn, P.~Edelaar, F.~J. Weissing, On the origin of species by
	natural and sexual selection, Science 326~(5960) (2009) 1704--1707.
	
	\bibitem{uyeda2009drift}
	J.~C. Uyeda, S.~J. Arnold, P.~A. Hohenlohe, L.~S. Mead, Drift promotes
	speciation by sexual selection, Evolution 63~(3) (2009) 583--594.
	
	\bibitem{m2012sexual}
	L.~K. M’Gonigle, R.~Mazzucco, S.~P. Otto, U.~Dieckmann, Sexual selection
	enables long-term coexistence despite ecological equivalence, Nature
	484~(7395) (2012) 506--509.
	
	\bibitem{nosil2012ecological}
	P.~Nosil, Ecological speciation, Oxford University Press, 2012.
	
	\bibitem{fierst2010genetic}
	J.~L. Fierst, T.~F. Hansen, Genetic architecture and postzygotic reproductive
	isolation: evolution of bateson--dobzhansky--muller incompatibilities in a
	polygenic model, Evolution 64~(3) (2010) 675--693.
	
	\bibitem{gourbiere2010species}
	S.~Gourbiere, J.~Mallet, Are species real? the shape of the species boundary
	with exponential failure, reinforcement, and the “missing snowball”,
	Evolution 64~(1) (2010) 1--24.
	
	\bibitem{fraisse2014genetics}
	C.~Fra{\"\i}sse, J.~Elderfield, J.~Welch, The genetics of speciation: are
	complex incompatibilities easier to evolve?, Journal of evolutionary biology
	27~(4) (2014) 688--699.
	
	\bibitem{yamaguchi2013first}
	R.~Yamaguchi, Y.~Iwasa, First passage time to allopatric speciation, Interface
	Focus 3~(6) (2013) 20130026.
	
	\bibitem{bank2012limits}
	C.~Bank, R.~B{\"u}rger, J.~Hermisson, The limits to parapatric speciation:
	Dobzhansky--muller incompatibilities in a continent--island model, Genetics
	191~(3) (2012) 845--863.
	
	\bibitem{burger2006conditions}
	R.~B{\"u}rger, K.~A. Schneider, M.~Willensdorfer, S.~Otto, The conditions for
	speciation through intraspecific competition, Evolution 60~(11) (2006)
	2185--2206.
	
	\bibitem{pennings2007analytically}
	P.~S. Pennings, M.~Kopp, G.~Mesz{\'e}na, U.~Dieckmann, J.~Hermisson, An
	analytically tractable model for competitive speciation, The American
	Naturalist 171~(1) (2007) E44--E71.
	
	\bibitem{bolnick2007sympatric}
	D.~I. Bolnick, B.~M. Fitzpatrick, Sympatric speciation: models and empirical
	evidence, Annu. Rev. Ecol. Evol. Syst. 38 (2007) 459--487.
	
	\bibitem{gavrilets2009adaptive}
	S.~Gavrilets, J.~B. Losos, Adaptive radiation: contrasting theory with data,
	Science 323~(5915) (2009) 732--737.
	
	\bibitem{de2017speciation}
	M.~A. de~Aguiar, Speciation in the derrida--higgs model with finite genomes and
	spatial populations, Journal of Physics A: Mathematical and Theoretical
	50~(8) (2017) 085602.
	
\end{thebibliography}

\end{document}